\documentclass[twoside,12pt]{amsart}

\usepackage{multicol}
\usepackage{amsmath}
\usepackage{amssymb}
\usepackage{mathrsfs}
\usepackage{graphicx}
\usepackage{booktabs} 
\usepackage{enumitem} 
\usepackage{abstract} 
\usepackage{hyperref} 
\usepackage{color}
\usepackage[normalem]{ulem}
\usepackage[numbers,square,sort&compress]{natbib}

\usepackage{a4wide}   
\usepackage{todonotes}
\usepackage{soul}

\title[Uncertainty in cardiac myofiber orientation and stiffnesses]{Uncertainty in cardiac myofiber orientation and stiffnesses dominate the variability of left ventricle deformation response} %

\author{
  Roc\'io Rodr\'iguez-Cantano
  \and
  Joakim Sundnes
  \and
  Marie E. Rognes
}
\date{}

\begin{document}
\maketitle

\begin{abstract}
  Computational cardiac modelling is a mature area of biomedical
  computing, and is currently evolving from a pure research tool to
  aiding in clinical decision making. Assessing the reliability of
  computational model predictions is a key factor for clinical use,
  and uncertainty quantification (UQ) and sensitivity analysis are
  important parts of such an assessment. In this study, we apply new
  methods for UQ in computational heart mechanics to study uncertainty
  both in material parameters characterizing global myocardial
  stiffness and in the local muscle fiber orientation that governs
  tissue anisotropy.  The uncertainty analysis is performed using the
  polynomial chaos expansion (PCE) method, which is a non-intrusive
  meta-modeling technique that surrogates the original computational
  model with a series of orthonormal polynomials over the random input
  parameter space. In addition, in order to study variability in the
  muscle fiber architecture, we model the uncertainty in orientation
  of the fiber field as an approximated random field using a truncated
  Karhunen-Lo\'eve expansion. The results from the UQ and sensitivity
  analysis identify clear differences in the impact of various
  material parameters on global output quantities. Furthermore, our
  analysis of random field variations in the fiber architecture
  demonstrate a substantial impact of fiber angle variations on the
  selected outputs, highlighting the need for accurate assignment of
  fiber orientation in computational heart mechanics models.
\end{abstract}

\section{Introduction}

Computational modelling of the heart is a powerful technique for
detailed investigations of cardiac behavior, and enables the study of
mechanisms and processes that are not directly accessible by
experimental methods. There is currently a drive towards adapting
these computational models to individual patient data, to aid in the
creation of individualized diagnosis, clinical decision support, and
treatment
planning~\citep{lopez2012,Chabiniok2015,Sack2016,Zhu2017,Biglino2017,Wolters2005,Land2015}. However,
this model adaptation presents a number of challenges related to the
lack of available data and the fact that measurable data, needed for
patient-specific model input parameters, are inherently subject to
measurement uncertainties or intrinsic biological variability. For
clinical use of models, it is therefore of crucial importance to
quantify how these uncertainties propagate through the computational
model to impact the output quantities of interest. Such assessment
should be performed with uncertainty propagation and uncertainty
quantification (UQ) techniques~\citep{Roy2011,Smith2013}, complemented
by sensitivity analysis (SA) to identify the most significant input
variables~\citep{Saltelli2008}.

In the particular case of cardiac ventricular mechanics studies,
individualized model adaption involves image-based construction of
computational geometries as well as tuning of material
parameters~\citep{Krishnamurthy2013,Young2009,Bai2015,Wang2009,Niederer2011,
  Bayer2012,Holzapfel2009,Wang2013}. Since the mechanical properties
of cardiac tissue are strongly anisotropic, the local material
behavior typically depends both on a set of material parameters and on
the local orientation of the cardiac muscle cells, typically referred
to as the fiber- and sheet orientation. The local tissue structure
can be determined with diffusion tensor magnetic resonance imaging
(DTMRI), but this technique is still limited to \textit{ex vivo}
experiments. Patient specific models have been created by projecting
\textit{ex vivo} DTMRI datasets onto patient-specific geometries
obtained from computed tomography (CT) or magnetic resonance imaging
(MRI)~\citep{Helm2005,Lombaert2012,Toussaint2013,Krishnamurthy2013,Nagler2013}.
However, even in the \textit{in vitro} case the accuracy of DTMRI is
$\pm$ 10 degrees~\citep{Reese1995,Scollan1998,Hsu1998}. While this
accuracy may be sufficient in the context of computational cardiac
electrophysiology~\citep{Vadakkumpadan2012}, local variations of this
order have been shown to introduce sizeable variations in myofiber
stresses~\citep{Geerts2003}.

Rule-based or atlas-based methods represent a convenient alternative
for assigning fiber- and sheet orientation in patient specific
models. For instance, the Laplace-Dirichlet Rule-Based (LDRB)
algorithm~\citep{Bayer2012} is based on atlas data, and assigns a
generic tissue architecture to image-based patient-specific
geometries. This method obviously neglects potential individual
variations in tissue structure, but provides a reasonable averaged
fiber/sheet orientation. Lombaert et al.~\citep{Lombaert2012} built the first
statistical atlas of the cardiac fiber architecture using human
datasets (10 \textit{ex vivo} hearts imaged with DTMRI), providing the
spatial distribution of fiber angles with their variability within the
healthy population. Their results showed that the helix angle of the
fibers varies globally from $-41 ^{\circ}$ ($\pm 26 ^{\circ}$) on the
epicardium to $-66 ^{\circ}$ ($\pm 15 ^{\circ}$) on the
endocardium. The reported variability includes both true variability of the
fiber structure and errors due to acquisition and image
registration. Similarly, Moll\'ero et al.~\citep{Mollero2015} estimated and represented
the uncertainty of cardiac fiber architecture originating from the
lack of data for a given patient using the mean and principal modes of
variations among a given population of healthy hearts.

In spite of the potential impact for clinical use of the models, there
are relatively few examples of proper UQ and SA for mechanical models
of the heart. Osnes and Sundnes~\citep{Osnes2012} and Hurtado et al.~\citep{Hurtado2017} 
studied the impact of uncertainty in material parameters, while
Puijmert et al.~\citep{Puijmert2017} investigated the sensitivity of a cardiac
mechanics model to changes in myofiber orientation over an average
angle of about 8$^\circ$.  An increase in total pump work of 11-19
$\%$ was found in three different geometries, revealing that
implementing an accurate fiber field is important for achieving the
correct model output. Sensitivity of cardiac models to the
myofiber orientation was also highlighted
in~\citep{Hassaballah2014,Wang2013,Nikou2016}.

One explanation for limited use of UQ and SA in cardiac modeling is
the computational expense of the involved models. A popular
statistical approach is the Monte Carlo (MC) method, but this method
typically requires a large number of model evaluations for converged
results. If the base model is a realistic computational model of
cardiac mechanics, the resulting computational cost will be
substantial. Techniques such as the quasi-Monte Carlo
(QMC)~\citep{Wang2001,Rubinstein2007} and the multilevel Monte Carlo
(MLMC)~\citep{Giles2015} methods can significantly improve the MC
convergence rate, but their application may be limited and technically
complex. Recently, alternative approaches, such as the use of surrogate
models~\citep{Sudret2017} to mimic the behaviour of the full model
while being inexpensive to evaluate, have been of particular
interest. One such technique is the \emph{polynomial chaos expansion}
method (PCE)~\citep{Wiener1938,Yang2017}, which has previously been used
in UQ analysis of cardiac mechanics and
electrophysiology~\citep{Osnes2012,Swenson2011}.

The purpose of the present work is to present a PCE based method for
UQ in cardiac mechanics models, and to perform an initial UQ and SA study
including both global myocardial material properties and local
variability of the microstructure orientation. The study of global
material parameters is similar to the UQ analysis
in~\citep{Osnes2012}, but using a more realistic computational model
and including a detailed SA of key input- and output variables. The UQ
considering local variations in microstructure orientation is, to our
knowledge, the first of its kind. In this case, the
input was treated as a random field, and modeled as a truncated
Karhunen-Lo\`eve expansion (KLE)~\citep{loeve1950} in order to reduce
the dimensionality of the random field representation. The former is
used as a basis to build a reduced-dimensionality representation of
the random field, essential to manage UQ analysis in extremely
high-dimensional problems.  Although the fiber arrangement exhibit a
typical gross architecture, as we mention above, there are local and
individual variations through the ventricular wall, as well as
uncertainty derived from noisy measurements that may affect the global
mechanical properties of the model. The results give insight into the
applicability of the truncated KLE method for representing noisy fiber
architecture fields, and to the impact of such variations on global
response quantities.
\
\section{Models and methods}

\label{sec:methods}

The overarching objective of this paper is to illustrate and evaluate
the impact of input data uncertainty on the mechanical response of the
heart. We introduce the forward model for the mechanical behaviour of
the left ventricle and its numerical approximation in
Section~\ref{sc:cm} below and describe our UQ techniques subsequently
in Section~\ref{sec:uq}

\subsection{Cardiac ventricular forward model}
\label{sc:cm}

\subsubsection{Governing equations}
\label{sc:ge}

Let $D \subset \mathbb{R}^3$ be the computational domain representing
the left ventricle. We consider the quasi-static and pressure-loaded
mechanical equilibrium problem over this domain: find the displacement
$\boldsymbol{u}: D \rightarrow \mathbb{R}^3$ such that
\begin{equation}
  \label{eq:fs}
  -\nabla \cdot (\boldsymbol{FS}) = 0  \quad  \text{ in } D,
\end{equation}
where $\boldsymbol{F}$ is the deformation gradient
i.e. $\boldsymbol{F} = \nabla \boldsymbol{u} + \boldsymbol{I}$, and
$\boldsymbol{S}$ is the second Piola-Kirchhoff stress tensor.
Boundary conditions for~\eqref{eq:fs} are described below.

We assume that the material is hyperelastic, and therefore that the
Piola-Kirchhoff stress tensor $\boldsymbol{S}$ is the derivative of a
strain energy density $\Psi = \Psi(\boldsymbol{E})$ with respect to
the Green-Lagrange strain tensor $\boldsymbol{E}$, defined as
\begin{equation}
  \boldsymbol{E}
  = \frac{1}{2}
  \left ( \boldsymbol{F}^T \boldsymbol{F} - \boldsymbol{I} \right ) .
\end{equation}
In particular, we
consider a transversely isotropic, hyperelastic and almost
incompressible material, and apply the widely used constitutive model
of Guccione et al.~\citep{Guccione1995}. This model is defined relative to three
mutually orthogonal vector fields: a fiber field $f : D \rightarrow
\mathbb{R}^3$, a fiber sheet field $s : D \rightarrow \mathbb{R}^3$
and a sheet normal field $n : D \rightarrow \mathbb{R}^3$. The
strain energy density is then defined as:
\begin{equation}
  \label{eq:sef}
  \Psi(\boldsymbol{E})  = \frac{1}{2} C (e^W -1) + K (J \ln J -J + 1)
\end{equation}
with
\begin{equation}
  \label{eq:def:W}
  W = b_{ff}E_{ff}^2 + b_{xx}(E_{ss}^2 + E_{nn}^2 + E_{sn}^2 + E_{ns}^2)
  + b_{fx}(E_{fn}^2 + E_{nf}^2 + E_{fs}^2 + E_{sf}^2).
\end{equation}
Here, $E_{ij}$ are components of the Green-Lagrange strain tensor in
the local fiber ($f$), fiber sheet ($s$), and sheet normal ($n$) axis,
i.e.~$E_{ij} = j \cdot \boldsymbol{E} i$ for directions $f$, $s$ and
$n$. Additionally, $J$ is the determinant of the deformation gradient,
and $C$, $K$, $b_{ff}$, $b_{xx}$ and $b_{fx}$ are material
parameters. In particular, $b_{ff}$ and $b_{xx}$ are parameters
governing the material stiffness in the fiber and cross-fiber
directions, respectively, $b_{fx}$ represents the shear stiffness in
planes parallel to the fibers, $K$ is the incompressibility factor of
the myocardial tissue, and $C$ enters as a multiplicative factor in
the strain energy function.

\subsubsection{Geometry, mesh and fiber orientations}
\label{ssc:gmfo}
\begin{center}
\begin{figure}
  \includegraphics[height=10cm, angle=90]{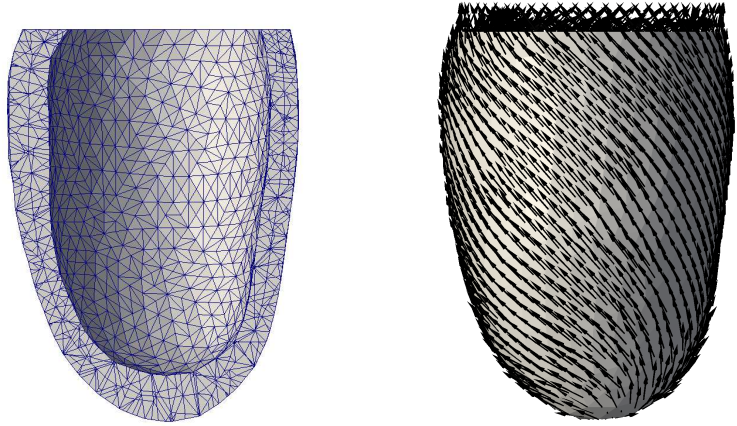}
\caption{Left: computational mesh of a human left ventricle
  wall. Right: Baseline fiber orientation field $f$ over the computational
  mesh. }
\label{fig:fig1}
\end{figure}
\end{center}

A computational mesh of the domain $D$ was generated from an
echocardiographic image of a left ventricle at the beginning of atrial
systole using the EchoPac software package (GE Healthcare Vingmed) and
Gmsh. We constructed a flat ventricular base by cutting the geometry
with a plane fit to the points on the base. The resulting linear
tetrahedral volumetric mesh of the left ventricle wall is shown in
Figure~\ref{fig:fig1} (left), counting $4\,507$ vertices and $18 \,
112$ cells.

As note above, the model~\eqref{eq:sef} assumes the availability of local
coordinate systems $f, s, n$ aligned with the local orientation of
muscle fibers. While the fiber orientation is not generally possible
to measure \emph{in vivo}, it is known that the fiber axes follow a
helical pathway as illustrated in Figure~\ref{fig:fig1} (right)
with a counter-clockwise rotation of the helix angle from epicardium
to endocardium~\citep{Streeter1969}. In view of this, we applied a
Laplace-Dirichlet Rule-Based (LDRB) algorithm~\citep{Bayer2012} to
generate realistic fiber-, fiber sheet- and sheet normal orientation
fields in our ventricular model. The LDRB method defines two main
angles to describe the local tissue structure. The fiber angle
$\alpha$ defines the orientation of the longitudinal fiber direction
relative to the circumferential direction, while $\beta$ is the angle
between the transverse fiber direction and the outward transmural axis
of the heart.

 Input parameters to the model are the values of these
angles on the endo- and epicardial surfaces, respectively:
$\alpha_{endo}$, $\alpha_{epi}$, $\beta_{endo}$ and
$\beta_{epi}$. Pointing ahead, in the present study we will both
consider these input angles as random variables, as previously done
in~\citep{Osnes2012}, and also apply a Karhunen-Lo\'eve expansion
(cf.~Section~\ref{sse:kle}) to study the impact of random variations
in the full fiber field.

\subsubsection{Boundary conditions}

Following e.g.~\citep{balaban2017high}, to constrain the displacement
at the base of the left ventricle boundary, we applied a Robin
boundary condition with a spring constant of $1$ kPa. Moreover, we let
the base of the left ventricle be clamped (zero displacement) in the
longitudinal direction. At the endocardium (inner) surface, we applied
a pressure of $2$ kPa, corresponding to the end-diastolic pressure, as a
normal stress boundary condition. At the epicardium (outer) surface,
we assumed zero normal stress.

\subsubsection{Numerical discretization}

To solve~\eqref{eq:fs} with the previously
described boundary conditions, we considered a finite element
discretization. The fiber-, fiber sheet-, and sheet normal orientation
fields were interpolated onto continuous piecewise linear vector
fields defined relative to the computational mesh, and we similarly
approximated the displacement field using continuous piecewise linear
vector fields. The nonlinear systems of equations were solved using
Newton's method and the resulting linear equations were solved
using a direct method. The endocardial pressure was applied
incrementally to improve the nonlinear convergence.

\subsection{Uncertainty quantification}
\label{sec:uq}

For brevity, in the presentation of the UQ techniques,
we will denote the finite element
discretization of the forward model described
by~\eqref{eq:fs} and associated boundary
conditions by $\mathscr{Y}$. In general, this forward model can be
viewed as a function, over the space $\boldsymbol{x} \in D$, mapping a
set of input parameters $\boldsymbol{\eta}$ to output values $Y$:
\begin{equation}
  \label{eq:forw}
  Y = \mathscr{Y}(\boldsymbol{x},\boldsymbol{\eta}).
\end{equation}
The mapping $\mathscr{Y}$ is deterministic, so that when evaluated on the
same $d$ input parameters ${\boldsymbol{\eta}} =
(\eta_1,\dots,\eta_d)$ it yields the same specific output values
$Y$.

We will consider both the case where each $\eta_i$
represents a (single) random variable and the case where some $\eta_i$
represent a random field. Concretely, $\boldsymbol{\eta}$ will
represent ventricular material parameters such as $C, K, b_{ff},
b_{fx}, b_{xx}$ and the input parameters of the fiber field model
$\alpha_{endo}, \alpha_{epi}, \beta_{endo}, \beta_{epi}$, or variables
associated with the uncertainty in the orientation field $f$.

A UQ analysis evaluates the impact in output $Y$ that results from the
uncertainty in the parameters $\boldsymbol{\eta}$, assuming a known
joint probability distribution $p_{\boldsymbol{\eta}}$ associated with
the input vector $\boldsymbol{\eta}$. The most popular technique for
UQ analysis is MC simulation, which involves the use of a sampling
method to draw a set of samples from the parameter space. Relevant
statistics of the output $Y$ is obtained by evaluating the
deterministic model~\eqref{eq:forw} on the sampling set. Although
simple and widely applicable, the MC technique converges slowly, and
typically requires a large number of evaluations of the forward model
$\mathscr{Y}$. In our case each evaluation involves solving a
non-linear finite element model, leading to a substantial
computational cost. We have therefore considered alternative
techniques to reduce the required number of $\mathscr{Y}$ evaluations.

\subsubsection{Polynomial Chaos Expansion}
\label{sse:pce}

The Polynomial Chaos expansion (PCE) method~\citep{Wiener1938} expands
the uncertain model outputs in a suitable series, which mimics the
behaviour of the forward model~\eqref{eq:forw} but is much cheaper to
evaluate. This series expansion can then be used to perform cheap UQ
and SA, using sampling techniques such as the QMC
method~\citep{Wang2001,Rubinstein2007}. In PCE, evaluations of the forward
model~\eqref{eq:forw} are required to build the series expansion, but
the number of required model evaluations is normally lower than for
standard sampling methods.

Assuming that the output of interest from~\eqref{eq:forw}
is a smooth function of $d$ random input parameters $\boldsymbol{\eta}
= (\eta_1,\dots,\eta_d)$, the PCE approximates $Y$ as a
function of $\boldsymbol{\eta}$ by a truncated polynomial expansion as
follows~\citep{Xiu2010}:
\begin{equation}
  \label{eq:PCE}
  Y(\boldsymbol{x}, \boldsymbol{\eta}) \approx
  \hat{Y}(\boldsymbol{x}, \boldsymbol{\eta}) =
  \sum_{i=1}^M c_i(\boldsymbol{x}) \Phi_i(\boldsymbol{\eta}).
\end{equation}
Here, $\{\Phi_i\}$ is a given multivariate orthogonal polynomial basis
for $\boldsymbol{\eta}$, $c_i(\boldsymbol{x})$ are the coefficients
that quantify the dependence of the model output on the parameters
$\boldsymbol{\eta}$, and $M$ is the total number of expansion terms.
This number is determined by the dimension $d$ of the random vector
$\boldsymbol{\eta}$ and the highest order $N$ of the polynomials
$\{\Phi_i\}$, more precisely $M = (N+d)!\, (N! \, d!)^{-1}$. The
deterministic functions $c_i(\boldsymbol{x})$ may be computed by the
point collocation method~\citep{Tatang1997}. Within this technique,
the unknown coefficients of the expansion are estimated by equating
model outputs and the corresponding polynomial chaos expansion at a
set of collocation points in the parameter space. For each output of
the model, a set of linear equations is formed with the coefficients
as the unknowns:
\begin{equation}
  \label{eq:pcm}
  \begin{pmatrix}
    \Phi_{1}(\boldsymbol{q}_1)\ & \cdots &\Phi_{M}(\boldsymbol{q}_1)    \\
    \vdots  &  \ddots & \vdots  \\
    \Phi_{1}(\boldsymbol{q}_{N_s}) &  \cdots & \Phi_{M}(\boldsymbol{q}_{N_s}) \\
  \end{pmatrix}
  \begin{pmatrix}
    c_{1}(\cdot)  \\
    \vdots  \\
    c_{M}(\cdot) \\
  \end{pmatrix}
  =
  \begin{pmatrix}
    \mathscr{Y}(\cdot, \boldsymbol{q}_1)  \\
    \vdots    \\
    \mathscr{Y}(\cdot, \boldsymbol{q}_{N_s})  \\
  \end{pmatrix}
\end{equation}
The collocation points $\{\boldsymbol{q}_1,\dots \boldsymbol{q}_{N_s}\}$
must be chosen in a way so that  the matrix~\eqref{eq:pcm} is
well-conditioned~\citep{Hosder2007}. This requirement allows for the
use of conventional QMC sampling methods~\citep{Dick2013} to select a
number of collocation points equal or greater than the number of
unknown coefficients $c_i(\boldsymbol{x})$~\citep{Tatang1997}.

Once the coefficients are determined and
$\hat{Y}(\boldsymbol{x},\boldsymbol{\eta})$ is built, the
last step in the PCE method for UQ is to propagate the uncertainties
through the simulator in order to estimate statistics of the response
quantities.  This last step is performed by MC simulations, in which
the model solver of \eqref{eq:forw} is substituted by the surrogate
$\hat{Y}$ as a cheaper alternative. It is important to note
that for PCE, the convergence depends on both the maximal order $N$ of
the polynomials $\{\Phi_i\}$ and the number of collocation points
$N_s$ selected to build
$\hat{Y}$. We return to
this point in Section~\ref{sec:results}. Typical statistical response
quantities include expected value ($\mu$), standard deviation
($\sigma$), prediction intervals and coefficient of
variation, in order to characterize the probability density function
(pdf) corresponding to each output quantity of
interest~\citep{Kendall1958}.

\subsubsection{QMC}

We have also applied Quasi-Monte Carlo
(QMC)~\citep{Wang2001,Rubinstein2007} simulations with Halton low-discrepancy
sampling sequences~\citep{Halton1960} to verify and validate the
results obtained by the PCE methods.

\subsubsection{Sobol sensitivity indices}
In addition to computing statistical properties of the output
probabilities, we perform
SA~\citep{Saltelli2002,Saltelli2008,Iooss2015} to quantify the
contribution of a particular input $\eta_i$, and of specific parameter
interactions, to the output variance. This analysis may be useful for
model personalization, for which \textit{input fixing} (identify
non-influential parameters to fix them within their uncertainty
domain) and \textit{input prioritization} (determination of which
factor(s), once fixed to its true value, leads on average to the
greatest reduction in the variance of an output) are important
goals. In this study we compute the total ($S_{i}^T$) and the main
($S_i$) variance-based Sobol sensitivity indices~\citep{Sobol2001},
which can be used for \textit{input fixing} and \textit{input
  prioritization}, respectively.

Specifically, the main sensitivity index $S_i$ is the proportion of
the total variance $\mathbb{V}$ of $Y$ that is expected to be reduced
if $\eta_i$ was fixed on its unknown true value. It can be computed
according to~\citep{Eck2015}:
\begin{equation}\label{eq:min}
 S_i = \frac{\mathbb{V}[\mathbb{E}[Y|\eta_i]]}{\mathbb{V}[Y]} ,
\end{equation}
where the index $i$ varies from 1 to the number of random inputs $d$,
and $\mathbb{E}$ is the expected value of the output quantity in
question $Y$.  Furthermore, the total sensitivity index
$S_{i}^T$, which represents the total variance due to both the direct
effect and all input interactions of $\eta_i$, is given
by~\citep{Eck2015}:
\begin{equation}\label{stotal}
 S_{i}^T = \frac{\mathbb{V}[Y] - \mathbb{V}[\mathbb{E}[Y|\boldsymbol{\eta}_{i*}]] }{\mathbb{V}[Y]}
\end{equation}
in which $\boldsymbol{\eta}_{i*}$ contains all uncertain inputs except
$\eta_i$.

\subsubsection{Karhunen-Lo\`eve expansion}
\label{sse:kle}
One of the key goals of this paper is to quantify and evaluate the
impact of uncertainty originating from the variability of the myofiber
orientation field $f$ cf.~\eqref{eq:def:W}. As a statistical model for
an input which address variability as function of space, it must be
described by a random field variable. In particular, in the following we will
consider a random myofiber orientation field as the sum of a random
field perturbation and a fiber field generated by the LDRB method
of Bayer et al.~\citep{Bayer2012}.
\begin{equation}{\label{eq:myoff}}
  f(\boldsymbol{x}, \theta) = f_{LDRB}(\boldsymbol{x}) +
  F(\boldsymbol{x}, \theta),
  \quad
  \boldsymbol{x} \in D,
\end{equation}
where $\theta \in \Omega$ denotes the dependency of $f$ on some random property.
To represent the random field
perturbation, we make use of a truncated Karhunen-Lo\'eve
expansion. Any second-order random field $F(\boldsymbol{x}, \theta)$
defined over $D \times \Omega$, with covariance function $C$ and
expected value $\bar{F}$, can be represented by the Karhunen-Lo\'eve
expansion~\citep{loeve1950,Ghanem1991}, also known as the proper
orthogonal decomposition, as the following infinite linear combination
of orthogonal functions:
\begin{equation}
  \label{eq:kl}
  F(\boldsymbol{x},\theta) = \bar{F}(\boldsymbol{x}) +
  \sum_{k =1}^{\infty} \eta_k(\theta)\sqrt{\lambda_k}\phi_k(\boldsymbol{x}).
\end{equation}
In~\eqref{eq:kl}, $\bar{F}(\boldsymbol{x})$ is the expected value of
the stochastic field at $\boldsymbol{x}$, $\{\eta_k(\theta)\}$
represents a set of uncorrelated random variables (if
$F(\boldsymbol{x},\theta)$ is assumed to be Gaussian then
$\{\eta_k(\theta)\}$ are also independent), 
and $\{\lambda_k$, $\phi_k(\boldsymbol{x})\}$ are
eigenvalues and eigenfunction pairs of the homogeneous Fredholm
integral equation over $D$:
\begin{equation}
  \label{eq:fi}
  \int_{D} C(\boldsymbol{y},\boldsymbol{x}) \, \phi_i(\boldsymbol{y}) \, {\rm{d}\boldsymbol{y}} =
  \lambda_i \phi_i(\boldsymbol{x}),
\end{equation}
using the covariance function $C(\boldsymbol{y},\boldsymbol{x})$ as
kernel~\citep{Ghanem1991}.

In practice, the infinite series in~\eqref{eq:kl} may be truncated
after the terms corresponding to the highest $n_{\rm KL}$ eigenvalues
$\{\lambda_k\}$:
\begin{equation}{\label{eq:nkl}}
  F(\boldsymbol{x},\theta) \approx \tilde{F}(\boldsymbol{x},\theta) = \bar{F}(\boldsymbol{x}) +
  \sum_{k =1}^{n_{\rm KL}} \eta_k(\theta)\sqrt{\lambda_k}\phi_k(\boldsymbol{x}).
\end{equation}
The number of terms $n_{\rm KL}$ depends on the decay of eigenvalues,
which in turn depends on the smoothness of the covariance function
$C$. If the eigenvalues $\{\lambda_k\}$ decay sufficiently fast and
$n_{\rm KL}$ is large enough, $\tilde{F}$ provides a suitable
approximation of $F$.

In this study, as we consider a random field perturbation to the
myofiber orientations, we assume $\bar{F}(\boldsymbol{x}) = 0$,
without loss of generality. Moreover, we have chosen the squared
exponential covariance structure ~\citep{Rasmussen2005} as the
covariance function;
\begin{equation}\label{eq:secf}
 C(\boldsymbol{x},\boldsymbol{y}) =
 \sigma_{\rm KL}^2 \exp( -\frac{|\boldsymbol{x}-\boldsymbol{y}|^2}{2l^2}) \hspace{1cm}
 \forall \, \boldsymbol{x},\boldsymbol{y} \in D .
\end{equation}
Here $\sigma_{\rm KL}^2$ is the field variance controlling the typical
amplitude of the random field, and $l$ is the correlation length that
defines the typical length-scale over which the field exhibits
significant correlations. Considering the lack of experimental data
from which to estimate the spatial uncertainty associated to the
myofiber orientation field, we consider this choice of correlation
function to be a sensible starting point for study. Finally, in this
study, the approach of truncation has been to examine the decay of the
eigenvalues $\{\lambda_k\}$ in~\eqref{eq:kl} and keep the first
$n_{\rm KL}$ eigenvalues $\{\phi_k(\boldsymbol{x})\}$ so that the
contributions from the remaining eigenvalues are negligible.

This reduction of dimensionality of the stochastic space, from
infinite to $n_{\rm KL}$, provides a parametric representation of the
random field $F(\boldsymbol{x},\theta)$ through $n_{\rm KL}$ random
variables. The uncertainty of the fiber field now stems from the
vector of parameters ${\boldsymbol{\eta}} = (\eta_1,\dots, \eta_{\rm
  KL})$, with $\{\eta_{k}\}$ the uncorrelated random variables defined
in~\eqref{eq:kl}.  Standard uncertainty propagation methods, like MC
or PCE, can be used then to predict the influence of the variability
of the myofiber orientation \eqref{eq:myoff} on our model. As an error
measure for the random field truncation \eqref{eq:nkl}, we have used
the error variance introduced by Betz et al.~\citep{Betz2014}. In particular,
$n_{\rm KL}$ has been selected ensuring that in more than the $92\%$
of the discretized points $\boldsymbol{x}$, the error variance is
lower than 0.05. In our experiments, $n_{\rm KL}$ range from $4$ to
$16$ depending on the correlation length $l$ in \eqref{eq:secf}.

\subsubsection{Computing Karhunen-Lo\'eve approximation}
Analytical solutions of the eigenvalue problem~\eqref{eq:fi} rarely
exist, so in general it has to be solved
numerically~\citep{Atkinson1997,Hackbusch1995}. For this purpose, we
consider the weak formulation (Galerkin projection) of the system of
equations~\eqref{eq:fi} on a discretization of the domain $D$. In
particular, assume that we have a mesh $\mathcal{T}_h$ of the fixed
domain $D$ with vertices (nodes) $x_1 \dots, x_n$. Take a continuous
piecewise linear basis $\{v_1, \dots, v_n\}$ defined relative to this
mesh, and consider the generalized eigenvalue
problem~\citep{Khoromskij2009}: find $\phi_k$ and $\lambda_k$ such
that
\begin{equation}\label{eq:tplmp}
 T \phi_k = \lambda_k M \phi_k,
\end{equation}
where $M$ is the mass matrix:
\begin{equation}
 M_{ij} = \int_D v_i(\boldsymbol{x}) v_j(\boldsymbol{x}) \, {\rm d \boldsymbol{x}},
\end{equation}
and
\begin{equation}
 T = M Q M
\end{equation}
with $Q_{ij} = C(\boldsymbol{x}_i, \boldsymbol{x}_j)$ the covariance
matrix that emerges from the discrete representation of the random
field with covariance kernel $C$.

It is important to note that while the mass matrix $M$ is symmetric
positive definite and may be sparse, $T$ is symmetric positive
semi-definite and dense.  Since $Q$ is dense, we applied a data sparse
technique to store it with the Hierarchical matrix
($\mathcal{H}$-matrix) format~\citep{Khoromskij2009,Saibaba2016}.
Consequently, the computational cost of matrix-vector products
involving $Q$ is reduced from $\mathcal{O}(n^2)$ to
$\mathcal{O}(n\log{}n)$, with $n$ the number of discretization
points. The $\mathcal{H}$-matrix technique is a hierarchical division
of a given matrix into rectangular blocks and further approximation of
these blocks by low-rank
matrices~\citep{Grasedyck2003,Hackbusch2000,Hackbusch1999}. In order
to compute the low-rank approximations, the Adaptive Cross
Approximation (ACA) algorithm~\citep{Bebendorf2003} was employed.

\subsubsection{Statistical properties of random input quantities}
\label{ssec:random:input}

\begin{center}
  \begin{table}[htp]
    \caption{Statistical properties of the input parameters in Model
      A: probability distribution ($\rho_{\eta_i}$), expected value
      ($\mu_{\eta_i}$) and standard deviation ($\sigma_{\eta_i}$).}
    \label{table:table1}
    \begin{tabular}{cllrr}
      \toprule
      Parameter & Unit & $\rho_{\eta_i}$ \ &$\mu_{\eta_i}$ & $\sigma_{\eta_i}$ \\
      \midrule
      $b_{ff}$ & & Normal & 6.6 & 0.99\\
      $b_{xx}$ & & Normal & 4.0 & 0.6\\
      $b_{fx}$ & & Normal & 2.6 & 0.39\\
      $K$  & kPa & Log-Normal & 10.0& 1.5\\
      $C$  & kPa & Log-Normal & 1.1& 0.165\\
      $\alpha_{endo}$ & degree & Normal  & 50.0& 7.5\\
      $\alpha_{epi}$  & degree & Normal  & 40.0&6.0\\
      $\beta_{endo}$  & degree & Normal  & 65.0&9.75\\
      $\beta_{epi}$   & degree & Normal  & 25.0&3.75\\
      \bottomrule
      \end{tabular}
  \end{table}
\end{center}

In this study, we introduce two different models of
uncertainty. First, we consider the material stiffnesses $b_{ff},
b_{xx}, b_{fx}$, the incompressibility parameter $K$ and the
weighting factor $C$ as uncertain (random) variables of prescribed
probability distributions. The statistical properties for these
material parameters were chosen as
in~\citep{Osnes2012}. Moreover, we similarly treat randomness in fiber
orientations as a direct function of the random input variables
$\alpha_{endo}$, $\alpha_{epi}$, $\beta_{endo}$ and $\beta_{epi}$ to
the LDRB algorithm. For these variables, we have assumed a normal
distribution with expected values following~\citep{Helm2005} and a
coefficient of variation equals to 0.15. The prescribed distributions,
expected values and standard deviations are listed in
Table~\ref{table:table1}, and we refer to this case as Model A.  All
parameters are treated as independent.

In the second model (Model B), we introduce uncertainty in the fiber
orientation field only by adding a Gaussian random field to the fiber
architecture generated by the LDRB algorithm. We thus introduce a
non-uniform perturbation in angle orientation of every fiber axis over
the computational geometry. The random perturbation field is
approximated via the truncated Karhunen-Lo\'eve expansion as
described in Section~\ref{sse:kle}. The properties of the random field
depend strongly on the selected correlation length. We have
considered three different correlation lengths: $l=3, 5$ or $10$ cm,
and two different standard deviations, $\sigma_{\rm KL} = 0.1$ and 0.5 
radians, respectively. In this second model (Model B), the five material 
parameters $C, K, b_{ff}, b_{fx}, b_{xx}$ and the angles $\alpha_{endo}, \alpha_{epi},
\beta_{endo}, \beta_{epi}$ are kept fixed at their mean value given by
Table~\ref{table:table1}.

\begin{figure*}
\begin{center}
    \includegraphics[width=0.8\linewidth]{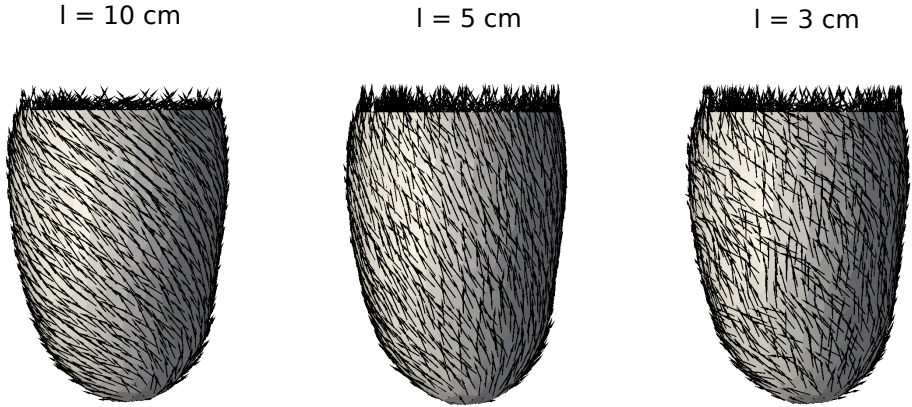} \par
  \caption{Samples of random fiber orientation
    fields (Model B), $f(\boldsymbol{x}, \theta)$ in \eqref{eq:myoff}, generated
    by a Gaussian random field with a standard deviation $\sigma_{\rm
      KL} = 0.5$ radians and correlation length equals $l=10$ cm
    (left), $l=5$ cm (middle) and $l=3$ cm (right).}
\label{fig:fig2}
\end{center}
\end{figure*}

Three samples of the different (total) random fiber orientation
fields, $f(\boldsymbol{x}, \theta)$ in \eqref{eq:myoff} assuming a 
standard deviation of 0.5 radians, are illustrated in Figure~\ref{fig:fig2}. 
Note that short correlation lengths in the random field generates 
strong fluctuations in the fiber architecture, while a higher value 
of $l$ implies that the random field approaches a random variable 
(i.e.~constant over the computational domain). The required number 
of terms $n_{\rm KL}$ in the Karhunen-Lo\'eve decomposition~\eqref{eq:nkl} 
varies from 4, 9, and 16 with decreasing $l$, so the more correlated 
the orientation field, the smaller the number of terms necessary to retain its
essential information in the truncated Karhunen-Lo\'eve expansion.

\subsubsection{Quantities of interest}

As quantities of interest (or target values) we have chosen global,
observable quantities: the volume of the inner cavity $Q_{c}$, the
lengthening of the apex $Q_{l}$ (difference between epicardial
and endocardial axial length), the change in wall thickness $Q_{t}$
(difference between outer and inner radius at base), and the total
wall volume $Q_{v}$. The reference values of these quantities of
interest (corresponding to the reference configuration of the
ventricular domain at zero endocardial pressure) are given in
Table~\ref{table:table2}.
\begin{center}
  \begin{table}[htp]
    \caption{Quantities of interest corresponding to the reference
      configuration of the ventricular domain.}
    \label{table:table2}
    \begin{tabular}{lr}
      \toprule
      Quantity of interest (unit) & Reference value \\
      \midrule
      Inner cavity volume $Q_c$ ($10^2$ $\times$ $\rm cm^3$) & 1.70  \\
      Apex lengthening $Q_l$    ($\rm cm$)                   & 1.11  \\
      Wall thickness $Q_t$      ($10^{-1}$ $\rm cm$)         & 6.99  \\
      Wall volume $Q_v$         ($10^2$ $\times$ $\rm cm^3$) &1.26   \\
      \bottomrule
    \end{tabular}
  \end{table}
\end{center}

\subsection{Implementation}

We used the Python interface to the FEniCS finite element
software~\citep{AlnaesBlechta2015a,Logg2012} to implement the forward
model described in Section~\ref{sc:cm}. The UQ analysis was performed
using the ChaosPy toolbox~\citep{Feinberg2015}, using the FEniCS
forward solver as a black box model. We also used FEniCS to assemble
the matrices $T$ and $M$ in~\eqref{eq:tplmp}. Finally, the dominant
eigenmodes of the eigenvalue problem~\eqref{eq:tplmp} (approximating
the eigenmodes of~\eqref{eq:fi}) were obtained using ARPACK accessed
via SciPy~\citep{Scipy2001}.

\section{Results}
\label{sec:results}

The main focus of this work is to quantify the impact of uncertainty
in local myofiber architecture on representative global response
quantities of interest. Prior to the main study focusing on model A
and B as described above, we present results from the calibration of
the surrogate PCE models.

\subsection{Surrogate model calibration and validation of statistical outputs}
The PCE model depends on the polynomial order $N$ ($\leq 3$) and number of
sampling points $N_s$ ($\leq 4\times M$) used to fit the surrogate model to the finite
element model. In order to choose these parameters, for each of the
experiments below, we conducted a series of experiments ultimately
choosing the $N$ and $N_s$ with the minimal mean-square error between
the surrogate and the forward (finite element) model outputs for a
new/different set of points in the parameter space.

Moreover, an extra convergence test has been performed comparing the
standard deviation of every response quantity obtained via this
validated single surrogate model with the same magnitude extracted
from a QMC simulation through the Halton low-discrepancy sampling
sequence. The results are included in Tables~\ref{table:table3} and
\ref{table:table4}--\ref{table:table5}. Overall, the 
non-intrusive PCE method was able to successfully generate a surrogate 
model for each quantity of interest specified in Table~\ref{table:table2}.

\subsection{Impact of input variable uncertainty}

\begin{center}
\begin{figure}
    \includegraphics[width=0.9\linewidth]{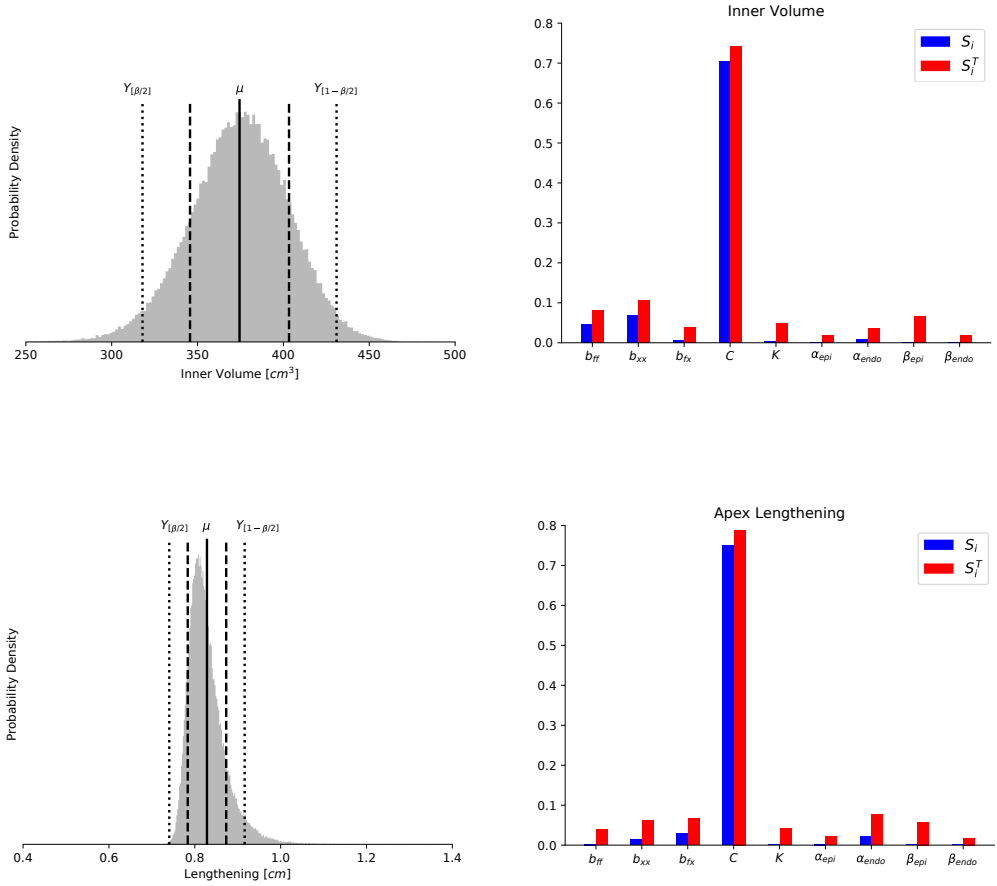}

\caption{Model A: Probability density (left column) of the inner
  cavity volume (top) and apex lengthening (bottom) obtained assuming
  material input parameters included in Table~\ref{table:table1}. In
  vertical bars, the mean (solid line), mean $\pm$ standard deviation
  respectively (dashed lines), and the limits of the 95$\%$ prediction
  interval (dotted lines) are shown. Main Sobol' index ($S_i$, blue)
  and total Sobol' index ($S_i^T$, red) are depicted in right column
  for all quantities of interest.}
\label{fig:fig3}
\end{figure}
\end{center}

\begin{center}
\begin{figure}
    \includegraphics[width=1\linewidth]{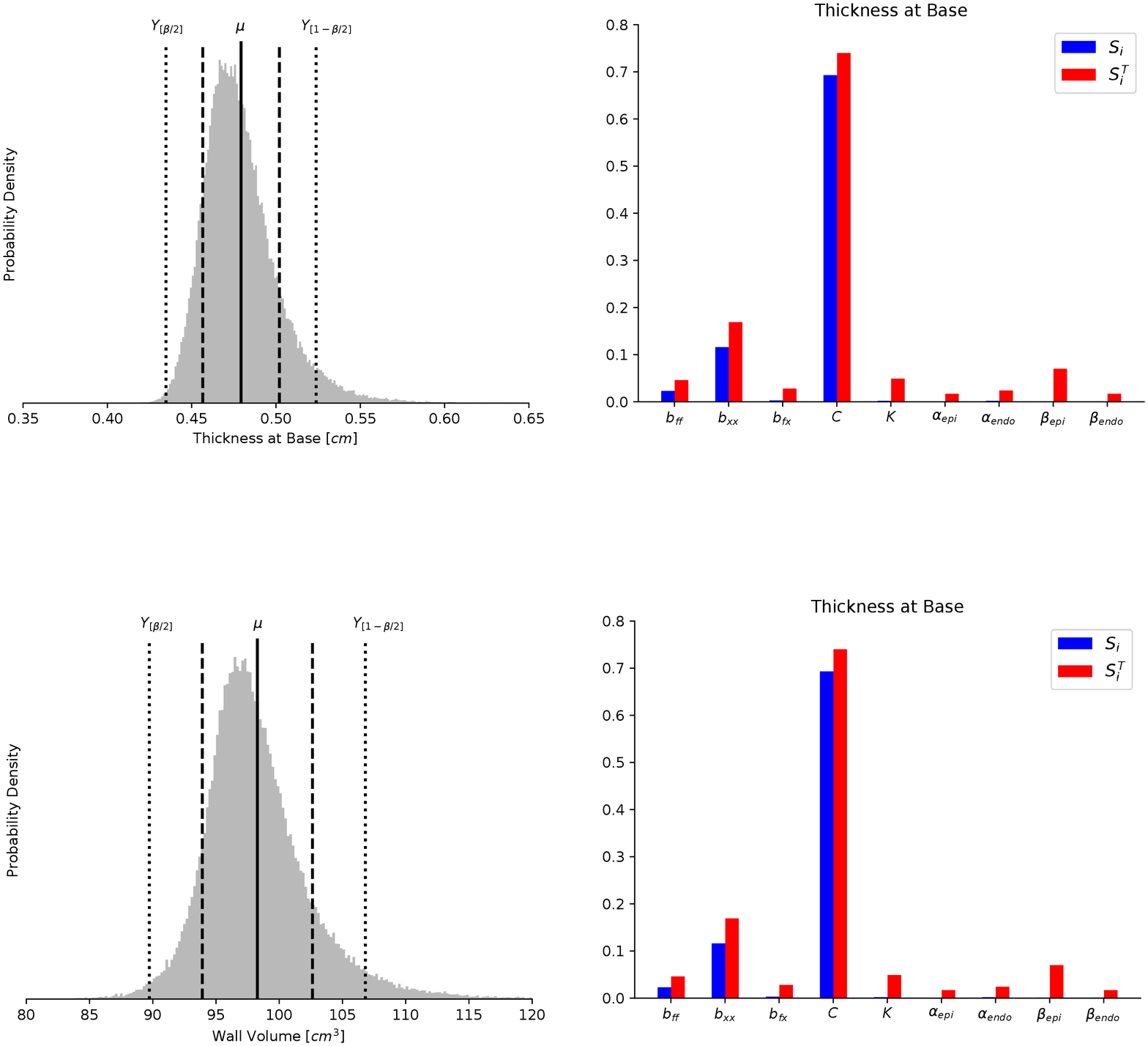}
  \caption{Model A: Probability density (left column) of the wall
    thickness at base (top) and wall volume (bottom) obtained assuming
    material input parameters included in Table.\ref{table:table1}.  In
    vertical bars, the mean (solid line), mean $\pm$ standard
    deviation respectively (dashed lines), and the limits of the
    95$\%$ prediction interval (dotted lines) are shown.  Main Sobol'
    index ($S_i$, blue) and total Sobol' index ($S_i^T$, red) are
    depicted in the right columns for all quantities of interest.}
\label{fig:fig4}
\end{figure}
\end{center}

We first consider a UQ analysis of Model A, in particular of the nine
model input random variables listed in Table~\ref{table:table1} and the
four output quantities of interest listed in
Table~\ref{table:table2}. We computed statistical properties of the
probability density functions associated with these output quantities,
including mean value $\mu$, standard deviation $\sigma$, coefficient
of variation $\sigma/\mu$ and the 95$\%$ prediction interval for each
output quantity. The resulting statistical quantities are listed in
Table~\ref{table:table3} and the output density functions are
depicted in Figures~\ref{fig:fig3}--\ref{fig:fig4} (left panels). We
observe that all coefficients of variation are at or below $0.08$,
with the largest coefficient of variation associated with the inner
cavity volume, and the smallest with the wall volume.

\begin{center}
  \begin{table}[htp]
    \caption{Model A: Statistical properties of the quantities of
      interest probability densities: expected value ($\mu$), standard
      deviation ($\sigma$), coefficient of variation (cov =
      $\sigma/\mu$), and prediction interval (PI$_{95}$) via PCE.
      Standard deviation extracted from QMC simulations
      is also included (QMC).}
    \label{table:table3}
    \begin{tabular}{ccccc}
      \toprule
      Quantity & $\mu$ & $\sigma$ (QMC) & cov  & PI$_{95}$ \\
      \midrule
      Inner volume   ($10^2$ $\times$  $\rm cm^3$)      & 3.81 &  0.30 (0.30) & 0.08 & [3.23, 4.39] \\
      Lengthening    ($\rm cm^3$)                       & 0.83 &  0.05 (0.05) & 0.06 & [0.73, 0.93] \\
      Wall thickness ($10^{-1}$ $\rm cm^3$)             & 4.79 &  0.28 (0.27) & 0.06 & [4.24, 5.34] \\
      Wall volume    ($10$ $\times$  $\rm cm^3$)        & 9.83 &  0.43 (0.42) & 0.04 & [8.90, 10.7] \\

      \bottomrule
    \end{tabular}
  \end{table}
\end{center}

For verification purposes, we also compared the resulting standard
variation values with values obtained using QMC directly (without the
use of the surrogate PCE model), also listed in
Table~\ref{table:table3}. We observe that the discrepancy in the
standard deviation between the PCE and the QMC simulations are less
than $2\%$ for all output quantities.

From Figures~\ref{fig:fig3}--\ref{fig:fig4}, we observe that output
density distributions display a high degree of symmetry (skewness $\sim$ 0),
though with a certain distortion to the right in the case of the apex
lengthening especially, but also for the thickness and wall volume, and
slightly negatively skewed data in the case of the inner cavity volume.

In addition to the statistical properties reported in Table~\ref{table:table3},
we computed the main and total Sobol indices with respect to the input
random variables for each output quantity. The indices are plotted in
Figure~\ref{fig:fig3}--\ref{fig:fig4} (right panels) in conjunction
with the respective output quantities. Clearly, the uncertainty in the
multiplicative factor $C$ has the highest main sensitivity index $S_C$
for all four output quantities, with $S_c \geq 0.6$ for all four
cases. More precisely, these sensitivity indices indicate that if $C$
was known and fixed to its true value, then the uncertainty in the
four output quantities $Q_c,Q_l,Q_t$, and $Q_v$
would be reduced by 70$\%$, 75$\%$, 69$\%$ and 61$\%$, respectively.

For the inner cavity volume, wall thickness, and wall volume, the
material stiffness parameters $b_{ff}$ and $b_{xx}$ have main
sensitivity index in the range $0.02-0.15$, with $b_{xx}$ having
higher index than $b_{ff}$ in the case of thickness at base, while
$b_{ff}$ have higher index than $b_{xx}$ in the case of wall volume
and inner cavity volume. In particular, $b_{xx}$ yields a main
sensitivity index greater than $0.05$ in the cases of inner cavity
volume and thickness and thus emerges as a key parameter for these
output quantities. Similarly, $b_{ff}$ emerges as a key parameter for
the wall volume. For these output quantities, the main sensitivity
index for the other variables ($b_{fx}$, $K$, $\alpha_{epi}$,
$\alpha_{endo}$, $\beta_{epi}$, $\beta_{endo}$) essentially
vanish. For the apex lengthening, we observe that the main sensitivity
indices associated with $b_{fx}$, $b_{xx}$, and $\alpha_{endo}$ are
small but non-vanishing, while the main sensitivity index associated
with other variables ($b_{ff}$, $K$, $\alpha_{epi}$, $\beta_{epi}$,
$\beta_{endo}$) essentially vanishes. Overall, the direct
contributions of the angles $\alpha_{epi}$, $\beta_{endo}$ and
$\beta_{epi}$ to the total variance of any output of interest are
negligible: the main sensitivity index of all three angles is close to
zero for for all the model outputs.

The total-order sensitivity indices identify the input variables that
may be fixed over their range of variability without affecting some
specific output variance. Indeed, these are those inputs corresponding
to $S_i^T \simeq 0$ for all the quantities of
interest~\citep{Saltelli2008,Smith2013}. We observe that the angles
$\alpha_{epi}$ and $\beta_{endo}$ satisfy this condition. These input
parameters have the smallest total sensitivity indices for all the
output quantities, and in particular, these inputs could be fixed in
subsequent model calibrations within their range of uncertainty
introducing only about $2\%$ of the current output variances. Finally,
the total and main sensitivity indices are of similar
value for all input parameters, which indicates no significant
high-order interaction between the model inputs.


\subsection{Impact of fiber field uncertainty}

\begin{center}
  \begin{figure}
    \includegraphics[width=0.5\linewidth]{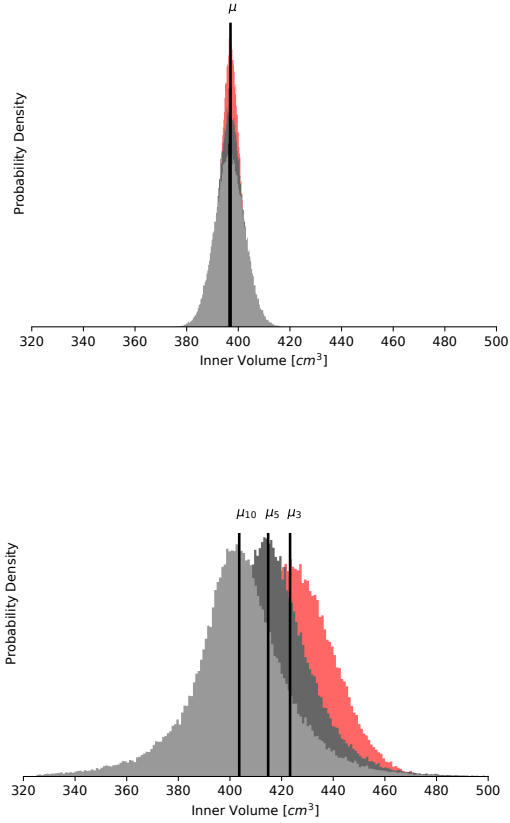} 
    \caption{Model B: Probability density functions for the inner
      volume cavity of the left ventricle assuming two Gaussian random 
      fields as uncertainty in the myofiber architecture: (top) 
      $\sigma_{\rm KL} = 0.1$ radians, (bottom) $\sigma_{\rm KL} = 0.5$ 
      radians. Colors correspond to different correlation lengths: red ($l=3$ cm), dark
      gray ($l=5$ cm) and light gray ($l=10$ cm).}
    \label{fig:fig5}
\end{figure}
\end{center}

\begin{center}
  \begin{figure}
    \includegraphics[width=0.5\linewidth]{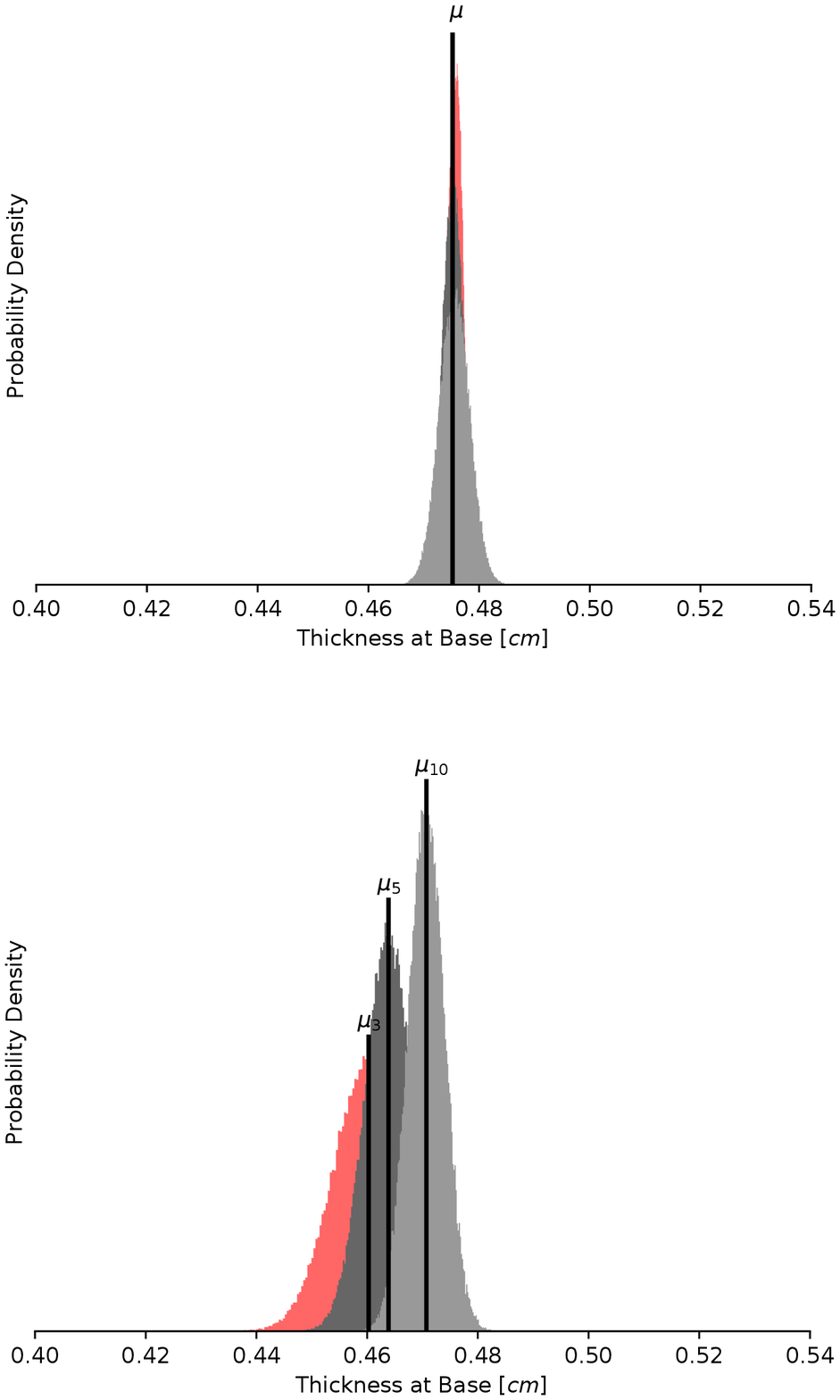} \\
    \caption{Model B: Probability density functions for the thickness at base
      assuming two Gaussian random fields as uncertainty in the myofiber 
      architecture: (top) $\sigma_{\rm KL} = 0.1$ radians, (bottom) 
      $\sigma_{\rm KL} = 0.5$ radians. Colors correspond to different 
      correlation lengths: red ($l=3$ cm), dark
      gray ($l=5$ cm) and light gray ($l=10$ cm).}
    \label{fig:fig6}
  \end{figure}
\end{center}

We next turn to consider a UQ analysis of Model B, where the local
fiber orientation is modeled as a Gaussian random field. We consider a
total of six cases; combining standard deviations $\sigma_{\rm KL}$ of
6.0 degrees (0.1 radians) and 28.6 degrees (0.5 radians) with
correlation lenghts 3, 5, and 10 cm. The chosen standard deviation 
values are based on fiber angle variabilities reported in the literature,
which include either measurement errors or a combination of
measurement error and biological variations. Samples of the resulting
random fields are illustrated in Figure \ref{fig:fig2}. 
For all six cases, we computed statistical properties of the probability 
density functions associated with these global response quantities, including 
the mean value $\mu$, standard deviation $\sigma$, coefficient of 
variation $\sigma/\mu$ and the 95$\%$ prediction interval for each 
output quantity listed in Table~\ref{table:table2}.
Statistical measures are presented in Tables~\ref{table:table4}--\ref{table:table5}, 
while the probability density functions are illustrated in Figures
~\ref{fig:fig5}--\ref{fig:fig6} for the inner volume and the 
wall thickness, as most relevant output quantities of interest.

For verification purposes, we also compared the resulting standard
variation values with values obtained using QMC directly (without the
use of the surrogate PCE model), also listed in 
Tables~\ref{table:table4}--\ref{table:table5}. We observe that the 
discrepancy in the standard deviation between the PCE and the QMC 
simulations is small, typically of the order $1-3\%$ for the range 
of output quantities and perturbation fields examined.

\begin{center}
  \begin{table}[htp]
    \caption{Model B: Statistical properties of the output quantities
      distributions: expected value ($\mu$), standard deviation
      ($\sigma$), coefficient of variation (cov = $\sigma/\mu$), and
      prediction interval (PI$_{95}$). Gaussian random fields with a standard 
      deviation $\sigma_{\rm KL} = 0.1$ radians and correlation length $l$ 
      equals to 3, 5 and 10 cm. Standard deviation extracted from QMC 
      simulations is also included (QMC).}
    \begin{tabular}{ccccccc}
      \toprule
      $l$ (cm) & Quantity & $\mu$ & $\sigma$ (QMC)  & cov ($\sigma/\mu$) & PI$_{95}$\\
      \midrule
       10 & Inner volume     ($10^2$ $\times$  $\rm cm^3$)     & 3.97  &  0.06 (0.06)  & 0.01  & [2.79,5.15]\\
          & Lengthening       ($\rm cm$)                       & 0.82  &  0.03 (0.03)  & 0.03  & [0.76,0.89] \\
          & Wall thickness    ($10^{-1}$ $\times$ $\rm cm$)    & 4.75  &  0.03 (0.03)  & 0.006 & [4.69,4.81] \\
          & Wall volume       ($10$ $\times$ $\rm cm^3$)       & 9.84  &  0.08 (0.09)  & 0.008 & [9.68,9.99]\\
      \midrule
      5  & Inner volume   ($10^2$ $\times$  $\rm cm^3$)       & 3.97  &  0.05 (0.05)   & 0.01  & [3.87,4.07]\\
         & Lengthening    ($\rm cm$)                          & 0.82  &  0.02 (0.02)   & 0.02  & [0.78,0.86]\\
         & Wall thickness ($10^{-1}$ $\times$ $\rm cm$)       & 4.75  &  0.02 (0.02)   & 0.004 & [4.71,4.79]\\
         & Wall volume    ($10$ $\times$ $\rm cm^3$)          & 9.84  &  0.07 (0.07)   & 0.007 & [9.70,9,98]\\

      \midrule
      3  & Inner volume      ($10^2$ $\times$  $\rm cm^3$) &  3.97  &  0.04  (0.04)    & 0.01   & [3.89,4.05]\\
         & Lengthening       ($\rm cm$)                    &  0.82  &  0.02  (0.02)    & 0.02   & [0.78,0.86]  \\
         & Wall thickness    ($10^{-1}$ $\times$ $\rm cm$) &  4.75  &  0.02  (0.02)    & 0.004  & [4.71,4.79]\\
         & Wall volume       ($10$ $\times$ $\rm cm^3$)    &  9.84  &  0.05  (0.05)    & 0.005  & [9.74,9.94]  \\

      \bottomrule
    \end{tabular}
    \label{table:table4}
  \end{table}
\end{center}
\begin{center}
  \begin{table}[htp]
    \caption{Model B: Statistical properties of the output quantities
      distributions: expected value ($\mu$), standard deviation
      ($\sigma$), coefficient of variation (cov = $\sigma/\mu$), and
      prediction interval (PI$_{95}$). Gaussian random fields with a standard 
      deviation $\sigma_{\rm KL} = 0.5$ radians and correlation length $l$ 
      equals to 3, 5 and 10 cm. Standard deviation
      extracted from QMC simulations is also included (QMC).}
    \begin{tabular}{ccccccc}
      \toprule
      $l$ (cm) & Quantity & $\mu$ & $\sigma$ (QMC)  & cov ($\sigma/\mu$) & PI$_{95}$\\
      \midrule
       10 & Inner volume     ($10^2$ $\times$  $\rm cm^3$)  & 4.03  &  0.18 (0.17)   & 0.04  & [3.70, 4.36]\\
         & Lengthening       ($\rm cm$)                     & 0.78  &  0.06 (0.06)   & 0.08  & [0.66, 0.90] \\
         & Wall thickness    ($10^{-1}$ $\times$ $\rm cm$)  & 4.70  &  0.05 (0.06)   & 0.01  & [4.60, 4.82] \\
         & Wall volume       ($10$ $\times$ $\rm cm^3$)     & 9.66  &  0.18 (0.17)   & 0.02  & [9.31, 10.0]\\
      \midrule
      5  & Inner volume   ($10^2$ $\times$  $\rm cm^3$)  & 4.15 &  0.18 (0.17)   & 0.04    & [3.80, 4.50]\\
         & Lengthening    ($\rm cm$)                     & 0.77 &  0.05 (0.05)   & 0.07    & [0.67, 0.87]\\
         & Wall thickness ($10^{-1}$ $\times$ $\rm cm$)  & 4.64 &  0.09 (0.09)   & 0.02    & [4.47, 4.81]\\
         & Wall volume    ($10$ $\times$ $\rm cm^3$)     & 9.44 &  0.21 (0.21)   & 0.02    & [9.04, 9.84]\\

      \midrule
      3  & Inner volume      ($10^2$ $\times$  $\rm cm^3$) &  4.24 &  0.18  (0.18)    & 0.04 & [3.89, 4.59]\\
         & Lengthening       ($\rm cm$)                    &  0.76 &  0.05  (0.05)    & 0.07 & [0.66, 0.86]  \\
         & Wall thickness    ($10^{-1}$ $\times$ $\rm cm$) &  4.60 &  0.06  (0.06)    & 0.02 & [4.48, 4.72]\\
         & Wall volume       ($10$ $\times$ $\rm cm^3$)    &  9.35 &  0.26  (0.28)    & 0.03 & [8.84, 9.86]  \\

      \bottomrule
    \end{tabular}
    \label{table:table5}
  \end{table}
\end{center}

Table~\ref{table:table4} shows the results of modeling the
fiber orientation as a Gaussian 
random field with a standard deviation of 6 degrees (0.1 radians). 
We see that all quantities of interest show a constant mean value,
independent of the correlation length. The coefficients of variation 
decreases slightly with decreasing correlation lengths for all 
quantities of interest except the inner volume, for which 
its coefficient of variation stays constant at 0.01 
across the correlation lengths investigated. 
Different and more interesting patterns 
are observed when the standard deviation is increased to 0.5 radians, 
as shown in Table~\ref{table:table5}. 
In this case cavity volume increases 
from 403 to 424 (cm$^3$) as the correlation length decreases 
from 10 to 3 (cm). The standard deviation stays approximately
constant, and the coefficient of variation is 0.04 for all correlation 
lengths. Turning now to the apex lengthening, we observe that 
this is the quantity of interest with the largest coefficient 
of variation, at $0.07-0.08$ for the correlation lengths 
examined, with slightly increasing coefficient of variation with 
increasing correlation length (Table~\ref{table:table5}). 
We also observe a slight decrease of the expected value with 
less correlation in the myofiber variability. Figure~\ref{fig:fig5}, 
shows density distributions for the cavity volume obtained 
with different perturbation fields. It can be seen that the degree of symmetry 
remains the same across the correlation lengths compared for both 
random fields under study. From kurtosis and skewness values 
(not shown), we confirm that the distributions associated to 
the inner volume, for both $\sigma_{\rm KL} = 0.1$ and 0.5 radians, 
can be considered as univariate normal distributions (absolute 
value of both skewness and kurtosis are within the range 
$\pm 1.96$~\citep{Gravetter2013,Trochim2006}). However, in the 
case of $\sigma_{\rm KL} = 0.5$ radians, as the correlation length $l$
increases, the distributions have heavier tails and 
sharper peak than the normal distribution, while as the correlation 
of the fiber noise decreases, the diastolic volume results 
are closer to a Gaussian curve.

From Table~\ref{table:table5}, 
we observe that the expected values of both the wall volume and wall
thickness decrease with decreasing correlation length, i.e. as the
uncertainty in the fiber field approaches a \emph{white noise}
field. The opposite trend holds in terms of the spread or relative
uncertainty for these two response quantities; the coefficient of
variation increases with increasing correlation length, though always
below 3$\%$. These findings are also contrary to the results mentioned 
above for a narrower width of the perturbation, for which 
the coefficient of variation diminishes as $l$ decreases.
Additionally, the skewness and kurtosis values for those two response 
quantities are close to zero (absolute value of both moments are 
within the range $\pm 1.96$~\citep{Gravetter2013,Trochim2006}) and 
thus wall volume and wall thickness distributions fit normal curves, 
as is also illustrated by Figure~\ref{fig:fig6} for the former quantity. 
It is interesting to note that in the particular 
case of correlation length $l=3$ cm and $\sigma_{\rm KL} = 0.1$ radians, 
we observe slightly negatively skewed data, with the left tail of the density 
distribution being longer and its mass concentrated on the right of the figure.

\section{Discussion}

The aim of this paper has been to analyze a computational model
describing the passive filling phase of the left ventricle using the
framework of uncertainty quantification. Our study quantifies the
impact of uncertainty in global material parameters, and, more
importantly, in measurements of local fiber orientations.  The
equations governing the passive
mechanical behaviour of the heart have been solved using the
finite-element software FEniCS~\citep{Logg2012}, and the implemented
uncertainty framework is based on polynomial chaos expansions
accessible via ChaosPy~\citep{Feinberg2015} and truncated
Karhunen-Lo\'eve expansions. This non-intrusive method allowed a
successful study of the impact of uncertainties, providing statistical
analysis through the probability densities of a set of global response
model outputs (inner cavity volume, apex lengthening, wall thickening,
and wall volume).

In our first simulation model, we identified the main uncertain input
parameters and characterized these by random variables obeying certain
specific probability distributions. The results clearly point at the
multiplicative factor $C$ as the parameter with the largest influence
on the variance in model outputs, and $K$, $\alpha_{epi}$,
$\beta_{endo}$ and $\beta_{epi}$ as the inputs with the lowest impact
on model response uncertainty.
Furthermore, the SA results indicate that
uncertainty in $C$ may  account for up to $75\%$ of the uncertainty
in the considered output quantities.
Our results suggest that the directional material stiffnesses,
both in fiber and cross-fiber directions, contribute less to overall
model output variance, but that these parameters are important for
wall volume and to some extent wall thickness. On the other hand,
our results from model A indicate that randomness in all angle
variables, except to some extent the angle at the endocardiac surface
($\alpha_{endo}$), contribute very little to the variance of all output
quantities of interest. Thus, even rather rough estimates of these
parameters would have little effect on the uncertainty in the output
predictions.

These findings may be compared to the results of~\citep{Osnes2012}
which also considered the influence of uncertainty in material
parameters in the mechanical response of the heart. First,
\citep{Osnes2012} identified the apex lengthening/ventricular elongation
as the output quantity most affected by uncertainty in input
parameters than the rest of the model outputs compared here. In
contrast, our results indicate that the model output with the largest
relative uncertainty (largest coefficient of
variation) is the variation of the inner cavity volume. Second, a
basic sensitivity analysis presented in~\citep{Osnes2012} revealed
that the inputs with the largest influence on the uncertainty of the
studied response quantities are $b_{xx}$ and $C$. Our results are in
partial agreement, as we found that $C$ and $b_{xx}$ are the
input parameters with the greatest influence on the output variance
only for the inner cavity and the wall volumes.
We note that these discrepancies between the present work and previous
results may be
due to differences in the considered mechanical model and
in the applied stochastic sensitivity analysis. In
particular, \citep{Osnes2012} considers an idealized and perfectly
symmetric geometrical model, which is likely to substantially impact
the results.

Modelling uncertainty in the input parameters for the LDRB algorithm
examines only one aspect of the influence of randomness in myocardial
fiber architecture. For a more thorough study, we therefore also
considered Gaussian perturbation fields of a base LDRB-generated fiber
orientation field, thus introducing local perturbations in fiber angle orientation
over the computational geometry.
Our results reveal that for moderate variability in fiber fields
($\sigma_{\rm KL} = 0.1$ radians), the impact on all output quantities
of interest is fairly low. The mean values stay constant 
independent of the correlation length of the field, and the
coefficients of variation are small and decrease slightly with
decreasing correlation length. For larger field variability
($\sigma_{\rm KL} = 0.5$ radians) the influence of the correlation in myofiber
uncertainty differs depending on the quantity of interest; for the
inner cavity volume the relative uncertainty does not change with the
correlation length of the perturbation field, while wall properties,
such as thickness or wall volume, experience a larger variation
relative to the mean as the correlation length decreases. In contrast,
our findings demonstrate the opposite behaviour for the lengthening
variation of the left ventricle at apex. Moreover, our results
indicate that the variability of the cardiac tissue in terms of fiber
arrangement has a greater influence on apex lengthening (coefficient
of variation up to 0.08) than any of the parameters considered in
model A (coefficient of variation 0.06). This is the only quantity of
interest where this is observed. Both for the apex lengthening and
inner cavity volume we found non-negligible coefficients of variation
for the variability in fiber orientation, independent of the
correlation length.

The apparent discrepancies between Model A and Model B have some
interesting implications. While most model outputs showed a very low
sensitivity to the global input parameters of the LDRB algorithm, the
experiment with large local local variations in fiber orientation
showed a large impact on the output quantities. These
results indicate that as long as a structured, helical arrangements of
fiber orientations is maintained, the precise angles of rotation are
not that important. On the other hand, any loss of the helical
structure, which is seen in Model B for low correlation lengths
(Figure \ref{fig:fig2} right panel), has a substantial impact on
the global mechanical properties of the ventricle. In real-world
applications, including patient specific simulations, the use of
rule-based assignment of fiber orientation will therefore tend to
exaggerate tissue organization and thereby the ventricular stiffness.
On the other hand, DTMRI based fiber fields capture both true tissue
variations and measurement noise, and are likely to underestemate the
inherent stiffness of the ventricle.

Future studies may target some of the limitations in this work as
discussed here. First, the input parameter uncertainty was modelled
using pre-specified normal/log-normal type distributions and
independent. For an even more realistic UQ analysis, one should
calibrate these probability distributions in accordance with
physiological or medical data (if available) via e.g.~Bayesian
inversion. Second, in order to quantify the variability of the fiber
architecture, while we here considered a truncated Karhunen-Lo\'eve
expansion, an alternative would be a Principal Component Analysis
(PCA). PCA may offer a more realistic quantification of the
variability of the fiber perturbation field by its mean and covariance
matrix sampled from a cardiac diffusion tensor imaging (DTI)
population distribution~\citep{Mollero2015}. Third, other extensions
of this study should include not just the filling phase of the heart
but also the active contraction of the muscle in the cardiac cycle, as
well as taking into account in the same model the uncertainty emerging
from both input material parameters and the fiber architecture.

Finally, an important limitation of the present study is that we only
consider the propagation of model parameter uncertainty through a
forward model of passive cardiac mechanics. In typical applications of
cardiac mechanics models, material parameters such as $C, b_{ff},
b_{xx},$, and $_{fx}$ are fitted to match data from patient recordings
or experiments. In this context the quantities considered as output
variables in the present study become input to a parameter estimation
problem~\citep{balaban2017high}. The results obtained in the present
study are valuable also in this context, since input variables with
high sensitivity indices will be most easily identifiable in an inverse
problem setting, while the variables with low sensitivity
are essentially non-observable. However, performing a proper UQ of this
parameter estimation problem,
quantifying how measurement error impact estimated parameters and in
turn model predictions, will be a highly relevant extension of the
present work.

\section{Conclusion}
We have performed a detailed UQ and sensitivity analysis of a
computational model of passive ventricular mechanics, using a PCE
method in combination with a
Karhunen-Lo\'eve expansion of stochastic field variables. The methods
were verified by comparing selected outputs with results of Quasi-Monte Carlo
simulations, confirming that the PCE approach gives an accurate and
computationally efficient representation of uncertainty propagation
through the cardiac mechanics model. The UQ and sensitivity analysis
can be concluded in two main findings. The first is that the
the multiplicative factor that scales the strain energy ($C$)
is the most sensitive parameter in the material law considered
here. The second is that while all considered model outputs are relatively
insensitive to the global endo- and epicardial helix angles, they are
highly sensitive to local variations and noise in the fiber
orientation.

\section{Acknowledgments}

This research is supported by the Nordic Council of Ministers through
Nordforsk grant \#74756 (AUQ-PDE), by the Research Council of Norway
through FRINATEK grant \#250731/F20 (Waterscape) and a Center of
Excellence grant to the Center for Biomedical Computing, and by NOTUR grant
NN9316K.



\end{document}